# Charge Transport and Conductance Switching of Redox-active Azulene Derivatives

Florian Schwarz,[a]+ Michael Koch,[b]+ Georg Kastlunger,[c]+ Heinz Berke,[b]* Robert Stadler,[c]* Koushik Venkatesan,[b]* and Emanuel Lörtscher[a]*

**Abstract:** Azulene (Az) is a non-alternating, aromatic hydrocarbon composed of a five-membered, electron-rich and a seven-membered, electron-poor ring; an electron distribution that provides intrinsic redox activity. By varying the attachment points of the two electrode-bridging substituents to the Az centre, the influence of the redox functionality on charge transport is evaluated. The conductance of the 1,3 Az derivative is at least one order of magnitude lower than those of the 2,6 Az and 4,7 Az derivatives, in agreement with density functional theory (DFT) calculations. In addition, only 1,3 Az exhibits pronounced nonlinear current-voltage characteristics with hysteresis, indicating a bias-dependent conductance switching. DFT identifies the LUMO to be nearest to the Fermi energy of the electrodes, but to be an active transport channel only in the case of the 2,6 and the 4,7 Az derivatives, whereas the 1,3 Az derivative uses the HOMO at low and the LUMO+1 at high bias. In return, the localized, weakly coupled LUMO of 1,3 Az creates a slow electron-hopping channel responsible for the voltage-induced switching due to the occupation of a single MO.

Controlling charge transport at nanometer length-scales is a challenge that requires knowledge of the underlying transport mechanisms. Functional molecular compounds featuring intrinsic charge-transfer control mechanisms are promising candidates for that task as the level of conductance can be defined deterministically by the chemical structure itself, involving discrete molecular orbitals (MO) with tuneable alignment to the Fermi energy of the electrodes, $E_F$, tailored conjugation across the molecular backbone, and variable anchoring schemes to the electrodes. In addition, conformational changes, chemical bond forming and breaking processes, quantum interference (QI), as well as charge- or spin-state alternations are molecular-intrinsic mechanisms, potentially addressable by an external stimulus to induce conductance switching or hysteresis.

Naphthalene ($C_{10}H_8$) is an alternating, bicyclic aromatic hydrocarbon, in which QI rules determine the transport properties through the various attachment sites. Symmetry-allowed and -forbidden pathways differ by more than one order of magnitude in conductance.[1] In azulene (Az), the non-alternating isomer of naphtalene, however, the question of the predictability of QI effects is currently under debate.[2] In addition, Az offers electronic features that allow transport to be controlled owing to its electron-donating and -accepting subsystems: For unsubstituted Az, the five-membered ring is electron-rich, whereas its edge-sharing seven-membered ring is electron-poor, inducing partial charge transfer to obey the Hückel 4*n*+2 rule for aromaticity. This results in a large dipole moment of μ = 0.8 - 1.08 D[3] and redox activity.

Here we report on the synthesis, the experimental transport characterization, and the theoretical modelling of three Az derivatives, in which the spatial orientation of the Az redox center was varied in respect to the transport path occurring between the two electrodes of a molecular junction (Scheme 1). These different orientations of the dipole moment and the electron-rich and -deficient moieties in respect to the field gradient are expected to cause large topological asymmetries for electron- or hole-transport regimes (a detailed theoretical study on the dipole moment can be found in the SI). Control of this variation is realized synthetically by attaching the two electrode-bridging substituents at variable nodes of the Az centre. In 1,3 Az (**1**), the transport-mediating side bridges are attached to the five-membered ring, whereas in 4,7 Az (**2**), they are attached to the seven-membered ring. In the remaining third rational variation, the anchor groups are attached to both rings at 2,6 positions (**3**). As anchors to gold (Au) electrodes, we use acetyl-protected sulphur terminals, whose protection groups are cleaved *in-situ* upon contacting the metal surface, forming the transport junctions Au-**1'**-Au, Au-**2'**-Au, and Au-**3'**-Au, respectively (Scheme 1). Because **2** and **3** are prone to polymerization if the 1 and 3 positions of the Az remain unblocked, ethyl groups were introduced. In all compounds, cyclic voltammograms exhibit well-defined and reversible oxidation waves, attributed to the oxidation of the five-membered ring of the central Az core (see the SI for the synthetic and $^{1}H$, $^{13}C$ NMR, IR, elemental, optical and electrochemical analyses).

The transport properties of individual compounds **1'** – **3'** were probed using a mechanically controllable break-junction system operated at low temperatures and ultra-high-vacuum conditions. Figure 1 A shows the entity of at least 300 current versus voltage, *I-V*, characteristics, taken at 50 K upon more than five dozens of opening and closing cycles[4] without any data selection. The colour code represents the relative occurrence of *I-V* data in these density plots. Because the measurement procedure aims at large mechanical manipulations to probe all possible configurations, also *I-V* characteristics close to direct Au-Au junctions are captured, still separated from Au-Au Quantum Point Contacts (QPC), as indicated by the white dotted lines (see SI for details). The statistically most probable current levels are determined by creating current histograms at arbitrary voltages (Figure 1 B), providing distinct accumulation peaks (with high current *I-V*s being neglected as identified by the transparent overlay). Based on the peaks, individual *I-V*s can be identified within the initial data sets (Figure 1 A) that represent the statistically most probable transport characteristics (light green curves in Figure 1 A). When taking the main peaks of the histograms, the order $I_{\mathrm{Au-1'-Au}} \leq I_{\mathrm{Au-2'-Au}} \sim I_{\mathrm{Au-3'-Au}}$ is found, with the relative difference being larger for small than for high bias because of the more pronounced nonlinear behaviour of **1′**.

[a] Dr. F. Schwarz, Dr. E. Lörtscher
Science and Technology Department
IBM Research - Zurich
Säumerstrasse 4, 8803 Rüschlikon, Switzerland
E-mail: eml@zurich.ibm.ch
[b] Dr. M. Koch, Prof. H. Berke, Dr. K. Venkatesan
Department of Chemistry
University of Zurich
E-mail: hberke@chem.uzh.ch, venkatesan.koushik@chem.uzh.ch
Winterthurerstrasse 190, 8057 Zürich, Switzerland
[c] Dr. G. Kastlunger, Dr. R. Stadler
Institute for Theoretical Physics
TU Wien – Vienna University of Technology
Wiedner Haupstrasse 8-10, Vienna 1040, Austria
E-mail: robert.stadler@tuwien.ac.at
[+] These authors contributed equally to this work.
[*] The authors thank SNF NRP 62 (406240-126142), the FWF (P27272), VSC (70671) and ÖAW, Springer Verlag and GÖCH.
Supporting information for this article is given via a link at the end of the document.

**Table 1.** Statistically most probable conductance, *G*, for the Au-**1'**-Au, Au-**2'**-Au, and Au-**3'**-Au junctions extracted from the entire data sets at biases of ±0.24 V, ±0.50 V, and ±1.00 V, respectively.

| | *G* @ ±0.24 V | *G* @ ±0.50 | *G* @ ±1.00 V |
|---|---|---|---|
| Au-**1'**-Au | $3.5 \times 10^{-6}$ $G_0$ | $7.1 \times 10^{-5}$ | $1.8 \times 10^{-3}$ $G_0$ |
| Au-**2'**-Au | $2.1 \times 10^{-3}$ $G_0$ | $4.9 \times 10^{-3}$ | $1.2 \times 10^{-2}$ $G_0$ |
| Au-**3'**-Au | $2.1 \times 10^{-3}$ $G_0$ | $7.5 \times 10^{-3}$ | $4.1 \times 10^{-2}$ $G_0$ |

In the literature, only single-molecule transport data for thiochroman-anchored Az derivatives measured at fixed bias (250 mV) under ambient conditions is reported.[2] There, the measured conductances overall were approximately one order of magnitude lower (than our 240 mV data) and reported to be a factor of four smaller for 1,3 Az and 2,6 Az than for 4,7 Az, whereas GW calculations with wide-band electrodes and the *ad hoc* assumption of $E_F$ to be at -1.5 eV suggested a conductance trend in the order $G_{1,3\ \mathrm{Az}} > G_{2,6\ \mathrm{Az}} \sim G_{4,7\ \mathrm{Az}}$.[2] The higher conductances in our experiments can be assigned to reduced electron-phonon interactions at low temperature and the covalently binding thiol anchors. The large disagreement between the two experimental data sets prompts an in-depth theoretical analysis that takes the applied field and realistic density of states for the electrodes into account.

Hence, we conducted nonequilibrium Green's function calculations within a DFT framework (NEGF-DFT). The transmission functions and *I-V* curves resulting from our calculations are shown in Figures 2 A and B, respectively, whereas in Table 2, the conductances derived from the *I-V* curves are listed. An agreement between theory and experimental data is found for the functional behaviour: **1'** shows distinct differences compared with **2'** and **3'**, with a nonlinear transport behaviour qualitatively reproduced by NEGF-DFT. The overall conductance is found to be higher by roughly one order of magnitude in theory because of the well-known gap problem of DFT and idealized junction configurations. Figures 2 C – 2 E show the evolution of MO eigenenergies with the bias.[5] In all three systems, the molecular LUMO (depicted in Figures 2 F – 2 H) is closest to $E_F$ and would therefore be expected to be the main contributor to the current. This is the case for **2'** and **3'**, where the molecular LUMO is delocalized over the entire molecule and strongly coupled to the electrodes. **1'**, however, exhibits a highly localized LUMO, which does not contribute to the current. In addition, a minimum is found in the transmission function near $E_F$ (Figure 2 A), explaining the relatively low conductance values for the Au-**1'**-Au system in both experiment and theory compared with Au-**2'**-Au and Au-**3'**-Au.

**Table 2.** Calculated conductances for Au-**1'**-Au, Au-**2'**-Au, and Au-**3'**-Au at biases of ±0.24 V, ±0.50 V, and ±1.00 V.

| | G @ | G @ | G @ |
|---|---|---|---|
| Au- | $1.6 \times 10^{-4}$ | $3.4 \times 10^{-4}$ | $3.1 \times 10^{-2}$ |
| Au- | $5.0 \times 10^{-2}$ | $8.1 \times 10^{-2}$ | $8.1 \times 10^{-2}$ |
| Au- | $3.6 \times 10^{-2}$ | $4.3 \times 10^{-2}$ | $2.7 \times 10^{-2}$ |

In addition to determining the single-molecule conductance statistically, our experimental approach allows the nonlinear transport properties to be determined as a function of bias when considering individual *I–V* characteristics (light green curves in Figure 1 A). As mentioned, **1'** reveals step-like features around 0.2 V, 0.5 V and 1.0 V, in contrast to **2'** and **3'** with rather smooth *I–V*s. Closer inspection (Figure 3 A) reveals a hysteretic behaviour in some of the *I–V*s (black and dark gray curves; ~ 5% of all curves), following a continuous transition from a lower to a higher conductive trace, with the intersection or crossing typically located around ±(0.85 - 1.00) V, in contrast to almost identical forward and reverse traces for the majority of curves (gray dotted curves).

Because of the chemical nature of the Az derivatives, switching mechanisms related to the spin degrees of freedom and the forming/breaking of particular bonds can be excluded, which leaves only two potential mechanisms for hysteresis: mechanical deformations allowing transitions between two stable conformers or a charging of the compound. We propose the latter option and show in the SI that metastable conformers do not exist from a DFT point of view. The observed behaviour of the *I-V* curves will be explained in the following by a two-channel model[6], in which the conductance is defined by a "fast" coherent tunnelling channel through delocalized MOs and a "slow" hopping channel through a localized MO, leading to conductance switching (and hysteresis) between a neutral and a charged state (Figure 3 D).

Of the Az derivatives investigated, only **1'** exhibits a transport-relevant MO that is localized (LUMO), whereas **2'** and **3'** do not reveal localized frontier orbitals at all, a fact that enables hopping-mediated charging only in the case of **1'**. *I-V* curves and transmission functions for neutral and charged compounds were simulated by using a two-channel model[6] combined with NEGF-DFT to describe the coherent tunnelling for the "fast" and DFT-based hopping rates according to Marcus theory[7] for the "slow" channel (Figure 3 B, C). The hopping path leads to a transient charging of compound **1’**, which results from a competition between the redox process and the experimental sweeping rate.[8,9] The agreement between the experimental (Figure 3 A) and simulated curves (Figure 3 B) is striking: whereas the neutral *I-V* curves still agree with the data shown in Figures 3 A (experiment) and 2 B (NEGF-DFT), the data for the charged compound (orange) reveals the same functional behaviour as the experimental ones (gray and black lines in Figure 3A) (additional theoretical and experimental data are given in the SI). The weakly coupled LUMO of **1’** therefore provides molecular-intrinsic, voltage-induced redox functionality within the solid-state junction because of the weak coupling of this particular MO to the electrodes. The appearance and functional variance of the hysteresis depend susceptibly on the coupling strength, whose amplitude is found to be a factor of 2 smaller for the **1’** LUMO than for the HOMO of an organometallic molybdenum derivative in Ref. 8. This fact could explain why hysteresis in the Au-**1’**-Au system is consistently observed, but less often (approx. 5 %) than in the organometallic system.[8] The voltage-induced two-channel molecular charging mechanisms in purely organic compounds is different to previously reported single-molecule conductance switching mechanisms[10] that were redox-based, but triggered by light[11] and electrochemical means[12] or due to conformational changes via the electric field.[13a,b]

In summary, the design and synthesis of three bis-alkynyl azulene (Az) derivatives coupled covalently via thioacetate groups to Au electrodes have been successfully accomplished. In the three compounds, the transport path through Az is varied systematically via variable attachment of the anchoring substituents over the 1,3, 4,7 and 2,6 positions. Statistics on the transport results reveal a current order following $I_{1,3\,\mathrm{Az}} < I_{4,7\,\mathrm{Az}} \sim I_{2,6\,\mathrm{Az}}$, consistent with DFT. In the unique case of 1,3 Az, *I–V* curves can be found that exhibit voltage-induced conductance switching and hysteresis. This behaviour is explained by a two-channel transport mechanism via a delocalized "fast" tunnelling channel for the conductance and a "slow" hopping channel for switching and hysteresis enabled by a very localized state close to the Fermi level. This study demonstrates that Az derivatives could be highly interesting building blocks for memory applications or neuromorphic devices in next-generation nanoelectronic applications.

## Experimental Section

Transport measurements were carried out using a Hewlett-Packard Parameter Analyzer 4156B and a statistical measurement approach.[4,14] DFT calculations used a NEGF formalism[15] for the description of electron transport at finite bias and the GPAW code[16] with a spacing of 0.18 Å for the real-space grid describing the potential energy term in the Hamiltonian, and a Perdew-Burke-Ernzerhof (PBE) parametrisation for the exchange-correlation (XC) functional. Owing to the high computational demands of finite bias calculations for these rather large junctions, we chose a linear combination of atomic orbitals (LCAO)[17] on a single zeta level with polarisation functions (SZP) for the basis set and one *k* point in the irreducible Brillouin zone in the scattering region.

## Acknowledgements

We are grateful to G. Puebla-Hellmann, V. Schmidt, and F. Evers for scientific discussions, and M. Tschudy, U. Drechsler and Ch. Rettner for technical assistance. We thank W. Riess, B. Michel and A. Curioni for continuous support.

**Keywords:** Azulene • Charge Transport • Redox-Activity • Molecular Switch • Hopping

# Schemes, Figures and Captions

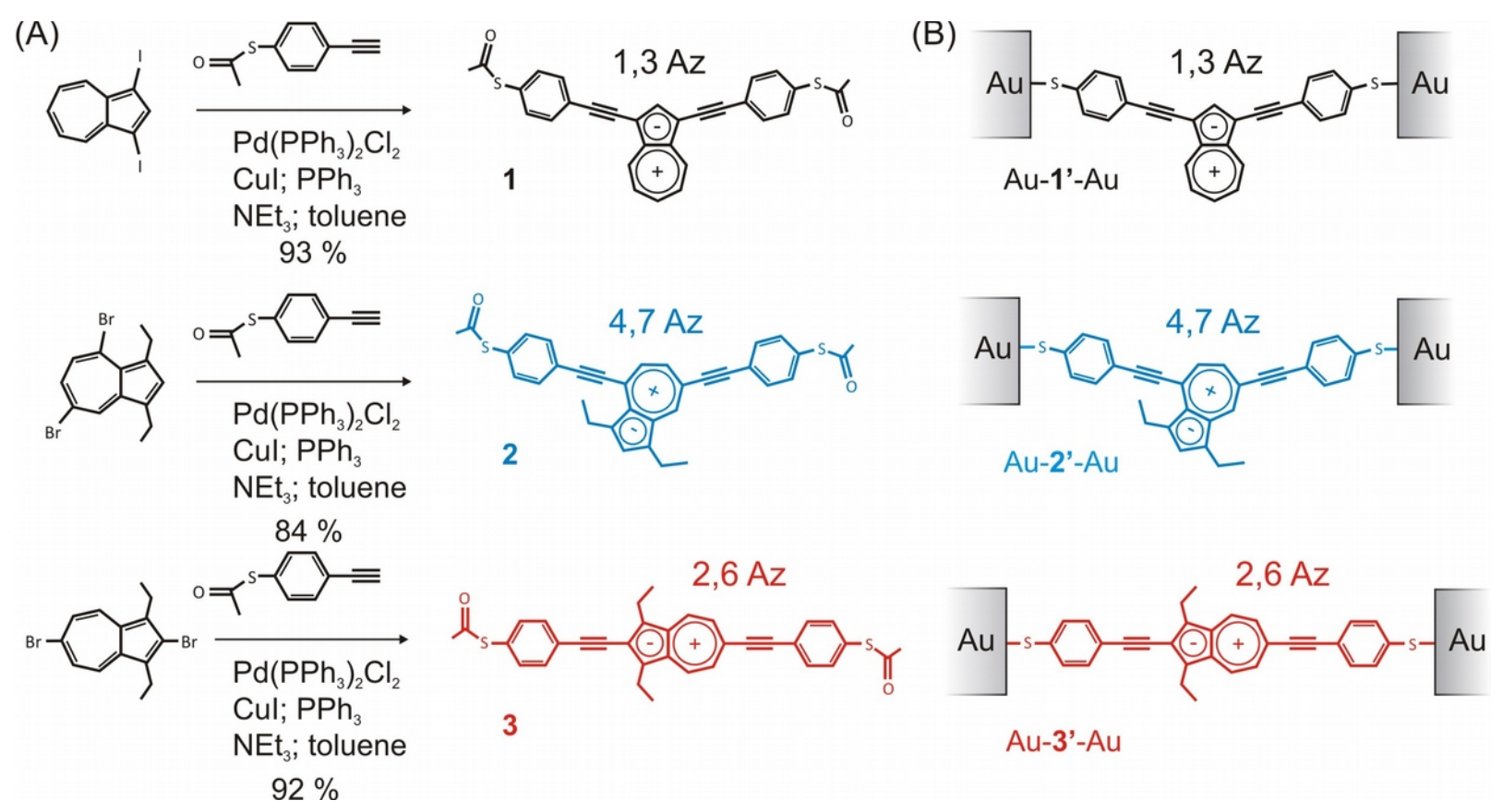

**Scheme 1.** (A) Simplified synthetic routes to produce the Az derivatives **1**, **2** and **3**, resulting in the transport junctions (B) Au-**1'**-Au, Au-**2'**-Au and Au-**3'**-Au, respectively. Details can be found in the SI.

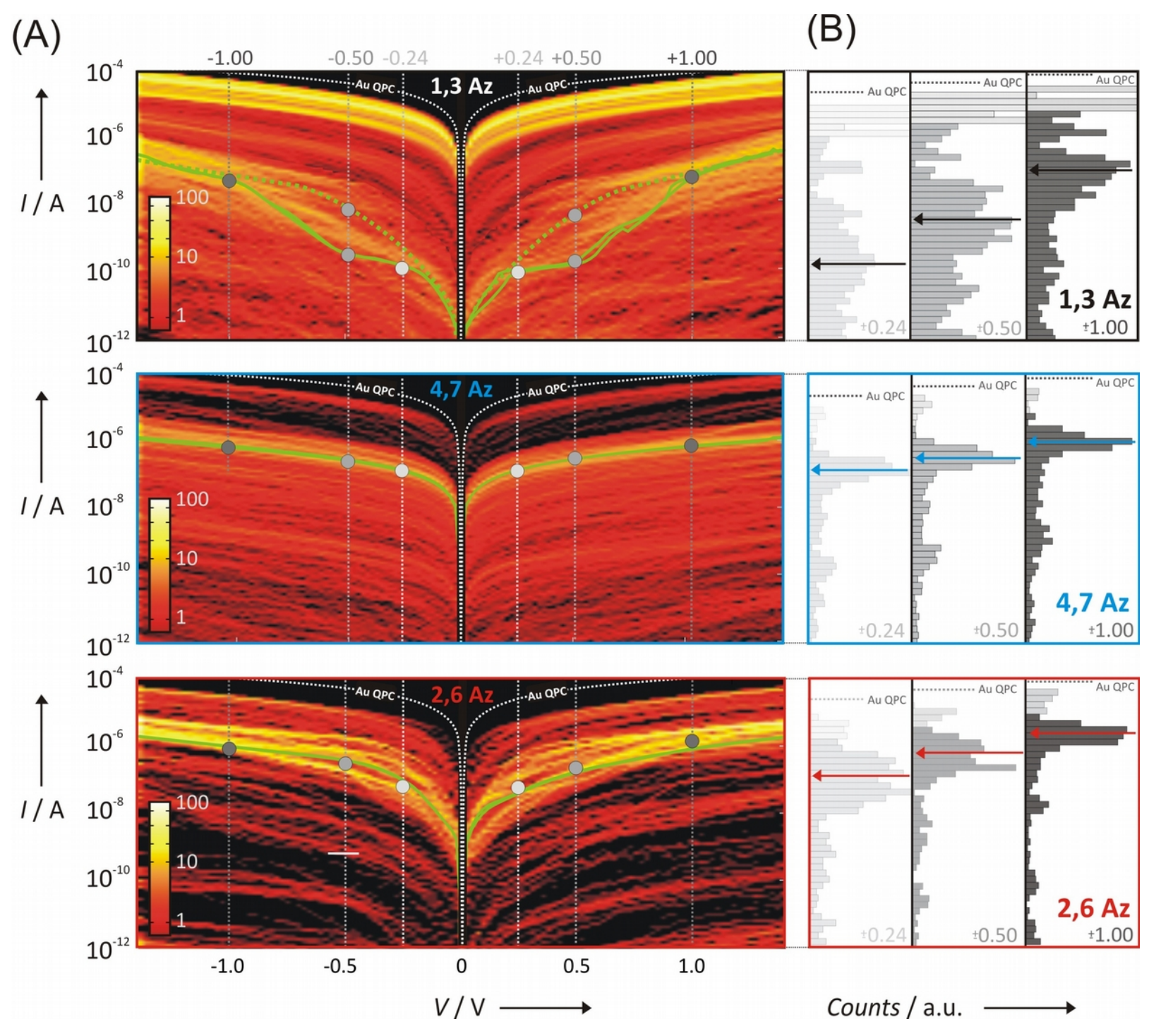

**Figure 1.** (A) *I-V* statistics over at least 300 opening and closing traces recorded for **1'**- **3'** at low temperatures of 50 K. (B) Histograms are created by extracting the current, *I*, from the entire data set (no data selection or filtering applied) at voltages, *V*, ±0.24, ±0.50 and ±1.0 V. The colour-coded lines indicate the most dominant single-molecule current peaks, whose values were converted into conductance values in Table 1.

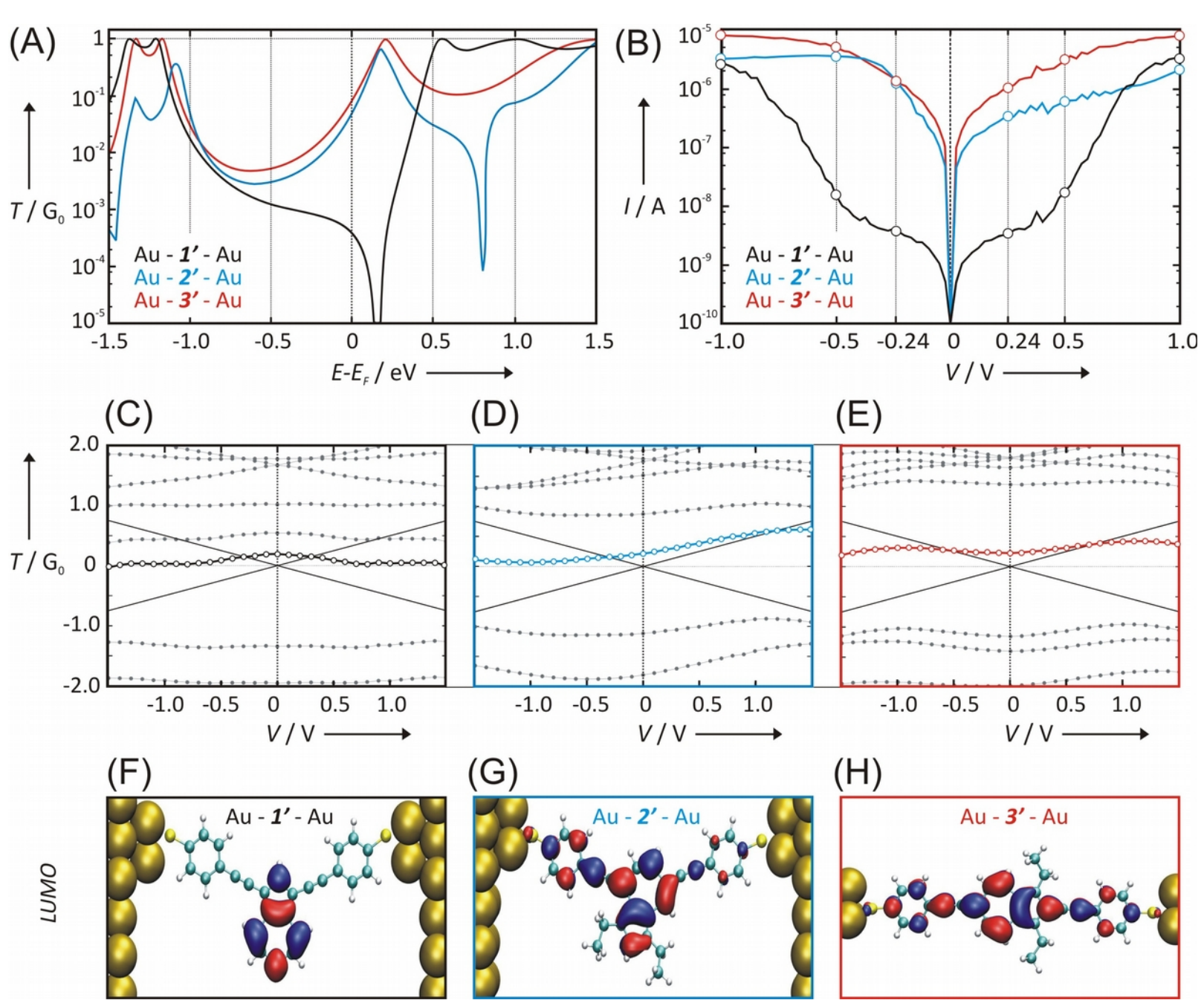

**Figure 2.** (A) Zero-bias transmission, *T*, as a function of energy, *E*-*E*$_F$, for Au–**1’**–Au (black), Au–**2’**–Au (blue), and Au–**3’**–Au (red) as calculated by NEGF-DFT. (B) Calculated bias-dependent *I*–*V* curves for the three uncharged compounds. (C-E) Bias-dependent evolution of the eigenergies of all MOs in the range ±2 eV around *E*$_F$ as obtained from a diagonalisation of the molecular subspace in the transport Hamiltonian. (F-H) Molecular LUMOs.

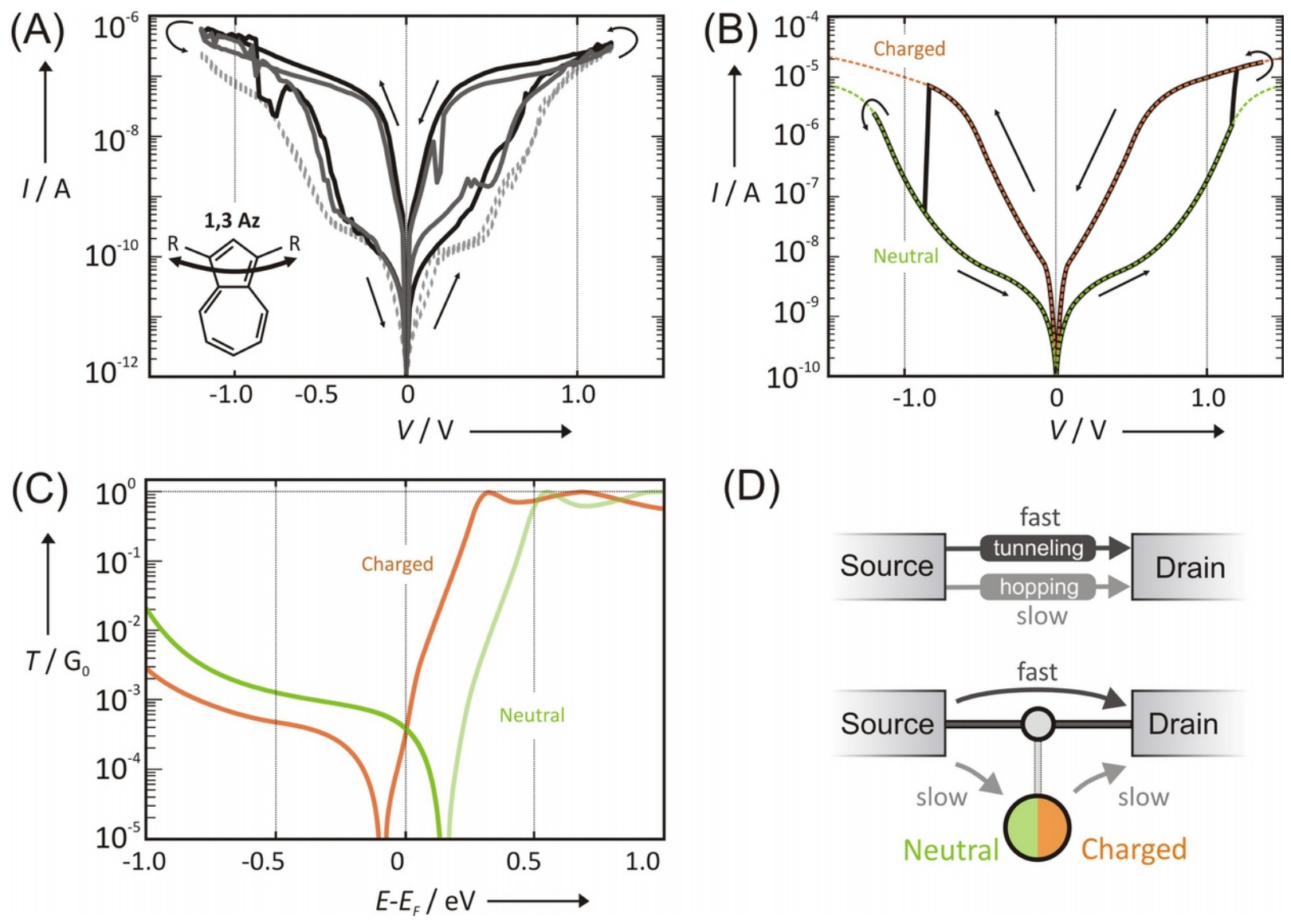

**Figure 3** (A) Hysteretic (black and dark gray curves) compared with non-hysteretic (dotted, light gray curves) *I*–*V* characteristics as found experimentally in selected cycles for **1’**. The hysteresis spans symmetrically in terms of bias, *V*, from 20 mV to ±(0.8 - 1.0) V. Arrows indicate the bias sweeping direction. (B) Simulated *I-V* characteristics (assuming a temperature of $T$ = 50 K, $V_{max}$ = 1.5 V, $nV$ = 60 and $\Delta t$ = 4 ms as well as hopping parameters for the slow channel of $\Delta G^0$ = -0.196 eV, $\lambda$ = 0.126 eV, $\Gamma = 2 \times 10^{-8}$, $dt$ = 0.004 ms, $n$ = 1000) revealing a hysteresis between the neutral (light green) and charged (orange) traces. (C) Transmission functions for the charged (orange) and neutral (light green) 1,3 Az compound as obtained from NEGF-DFT. (D) Simplified transport mechanism with a fast, coherent tunnelling and a slow, hopping channel.

By varying the attachment points to the Azulene center, the influence of the redox functionality on charge transport is evaluated. Among the three substituent patterns, only the 1,3 Az derivative revealed non-linear and hysteretic transport behaviour. Its weakly coupled LUMO is identified by DFT to be chargeable, leading to a transport mechanism involving additionally a slow electron-hopping channel responsible for the switching due to

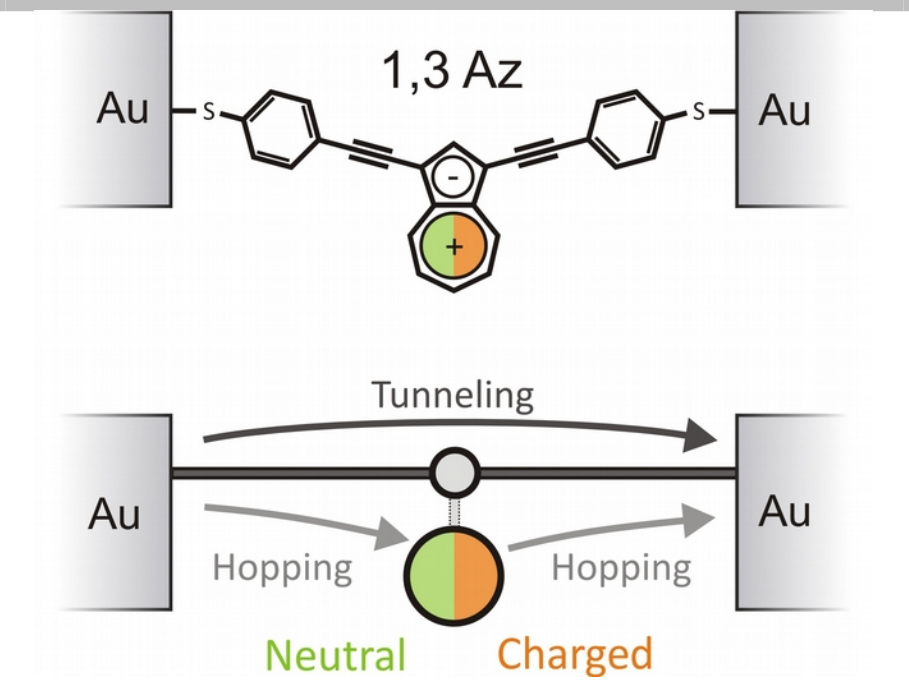

*Florian Schwarz, Michael Koch,Georg Kastlunger, Heinz Berke, Robert Stadler, Koushik Venkatesan, and Emanuel Lörtscher*

**Charge Transport and Conductance Switching of Redox-active Azulene Derivatives**